\documentclass[aps,pre,floatfix,twocolumn]{revtex4}
\usepackage{graphicx} \usepackage{amsmath} \usepackage{amssymb}\usepackage{color}
\newcommand{\comment}[1]{}
\newcommand{\BEQ}{\begin{equation}}
\newcommand{\EEQ}{\end{equation}}
\newcommand{\BEA}{\begin{eqnarray}}
\newcommand{\EEA}{\end{eqnarray}}

\begin{document}
\draft
\title{Comment on ``Cooling by Heating: Refrigerator Powered by Photons"}
\author{Armen E. Allahverdyan$^{1,4)}$, Karen V. Hovhannisyan$^{2)}$ and Guenter Mahler$^{3)}$}

\address{$^{1)}$Laboratoire de Physique Statistique et Syst\`emes Complexes,
ISMANS, 44 ave. Bartholdi, 72000 Le Mans, France\\
$^{2)}$ICFO-Institut de Ciencies Fotoniques, 08860 Castelldefels (Barcelona), Spain\\
$^{3)}$Universitat Stuttgart, 1. Institut für Theoretische Physik,
Pfaffenwaldring 57, 70550 Stuttgart, Germany\\
$^{4)}$Yerevan Physics Institute, Alikhanian Brothers Street 2, Yerevan 375036, Armenia }

\begin{abstract} 
Results obtained recently in Phys. Rev. Lett. {\bf 108}, 120603 (2012) by Cleuren et al.
apparently contradict to the third law of thermodynamics.  We discuss a
scenario for resolving this contradiction, and show that this scenario
is pertinent for clarifying the general message of the third law. 
\end{abstract}




\maketitle

{\bf 1.}
A recent interesting letter by Cleuren et al. presents a model for a
refrigerator operating between two thermal baths at temperatures $T_c$
and $T_h$, $T_c<T_h$ \cite{1}. Among other findings, Cleuren et al.
report on a low-$T_c$ regime, where the heat $\dot{Q}_c$ taken per time-unit
from the low-temperature bath scales as 
\BEA
\label{1}
\dot{Q}_c \sim T_c.
\EEA 
If (\ref{1}) is assumed to hold for all temperatures down to $T_c\to 0$,
it would invalidate (for this model example) the third law of
thermodynamics \cite{2}. This law states that the rate of cooling of the
low-temperature bath goes to zero. The challenge of saving the third law
amounts to showing why (\ref{1}) cannot hold down to very low
temperatures $T_c$.  A proposal on why this might be the case
for the model studied in \cite{1} was recently made by
Levy et al. \cite{2}. 

{\bf 2.} The purpose of the present note is to explain why to our
opinion the proposal by Levy et al. cannot be accepted as a fundamental
reason for saving the third law, and then make our own attempt of
arguing against the validity of (\ref{1}) at very low temperatures.  We
stress that we do not make any claim on the invalidity of the results by
\cite{1} within their model. It is the applicability of the model for
low temperatures that is questioned. 

{\bf 3.} The contradiction between (\ref{1}) and the third law is
deduced via routine thermodynamic considerations: {\it (i)} the
low-temperature thermal bath stays in overal equilibrium despite of its
interaction with the refrigerator. Hence it will respond to the
refrigerator by lowering its temperature.  {\it (ii)} The rate at which
its temperature $\dot{T}_c$ is lowered can be evaluated within the
linear response, since the energy taken out due to refrigeration is much
smaller than the energy of the bath. Hence \cite{2}
\BEA
\dot{T}_c=\frac{\dot{Q}_c}{c(T_c)},
\EEA
and taking into account that for $T_c\to 0$ the constant-volume heat
capacity $c(T_c)$ behaves at least as $c(T_c)\sim T_c$ for reasonable
thermal baths (including the electron bath studied in \cite{1}), one
concludes that $\dot{T}_c$ will be at least constant for $T_c\to 0$,
which contradicts to the third law.

{\bf 4.} To save the third law, Levy et al. propose that the Hamiltonian
of the refrigerator employed in \cite{1} is supplemented by another,
physically well motivated term that invalidates (\ref{1}) for a low
$T_c$ \cite{2}. This salvation of the third law is not satisfactory for
the following reason. 

Any model of refrigerator must describe its specific function. This
description necessarily involves taking ``limits", i.e.
letting certain parameters in the respective Hamiltonian go to zero. 
We distinguish two types of such asymptotic behavior:

{\it 1.} ``Circumstantial limits" strengthen the functional charateristics of the model. 
Applying such limits is a natural desire of
building better devices. Indeed, good devices do have rather special
Hamiltonians. As the evolution of room refrigerators shows, even when
their theoretical operating principles are clear, it still takes many
years and substantial engineering efforts to built good devices,
precisely because many unwanted terms in their Hamiltonians are to be
eliminated. 

{\it 2.} ``Dysfunctional" limits would suppress the desired function of
the device (asymptotically). 

Now Levy et al. based their arguments on an circumstantial limit as a
potential reason of violating the third law.  It may be difficult to
implement this limit in practice, but nothing in the analysis by Levy et
al. shows that the term they propose cannot {\it in principle} be made
as small as desired. This viewpoint on the salvation of the third law
would suggest that this law is not fundamental, but it holds due to
imperfections present in the Hamiltonian. 

In contrast, we are going to argue that the violation of the third law
by Cleuren et al. relates to a dysfunctional limit: if it is
applied down to very low temperatures, the basic functional
characteristics of the model (its power of refrigeration) will be
harmed.

{\bf 5.} Now we explain why the weak-coupling master-equation-based
refrigerator model by Cleuren et al. gets limited at low $T_c$. 
An essential feature of such Markov models is the detailed balance with respect to 
each thermal bath. Due to this, for $T_c=T_h=T$ (equal
temperature baths) the refrigerator density matrix $\rho$ has the Gibbs
form $\rho\propto e^{-H/T} $, where $H$ is the refrigerator's
Hamiltonian. For $T\to 0$ this predicts that the refrigerator itself
will be in its pure ground state. Such a conclusion is impossible for a
system permanently coupled to a thermal bath, provided that both the
system-bath interaction Hamiltonian and its commutator with the full
Hamiltonian stay finite (non-zero); see e.g. \cite{armen}. The model
studied in \cite{1} belongs to this class.

The usual way of understanding the low-temperature limit of the Gibbs
density matrix for an open system is to assume that simultaneously with
$T\to 0$ the coupling to the bath is made progressively smaller.
However, for the present situation this limiting process for justifying
the Gibbs density matrix down to low temperatures does not work, since
any refrigerator should have a finite coupling to the baths for ensuring
a finite power of its operation. While the argument strictly speaking
applies only for $T_c=T_h=T$, it should be clear that there are
low-$T_c$ validity limits of the weak-coupling master equation also for
$T_c<T_h$. Hence if the weak-coupling master equation is forced to apply
for all $T_c\to 0$, its coupling to the low-temperature bath should be
made progressively weaker nullifying $\dot{Q}_c/T_c$ in (\ref{1}). 

{\bf 6.} The above argument on the inapplicability of the usual Markov
models at low temperatures suggests that the analysis of the
low-temperature refrigeration will certainly benefit from being carried
out in a set-up, where the refrigerator bath interaction is treated
exactly, without any assumption on progressively weak refrigerator-bath
interactions. Now one should and can ensure that the power of
refrigeration stays finite down to $T_c=0$. Such a model-dependent
analysis has been carried out recently showing that albeits the regime
(\ref{1}) is reproduced by sufficiently good devices at moderate and low
values of $T_c$, it is broken down for very low temperatures holding the
third law \cite{3}. 

Our conclusion is that a proposition like (\ref{1}) should always be
supplemented by demanding that the power of refrigeration stays finite.
Then presumably it cannot hold down to very low temperatures, as the
model studied in \cite{3} shows. If such a proposition could be shown to
hold down to $T_c=0$ with only circumstantial limits involved, it would
constitute a a ``real" violation of the third law.

\end{document}